# Privacy-Preserving Distributed Learning Framework for 6G Telecom Ecosystems

Pooyan Safari
*Photonic Networks and Systems*
*Fraunhofer HHI*
Berlin, Germany
pooyan.safari@hhi.fraunhofer.de

Behnam Shariati
*Photonic Networks and Systems*
*Fraunhofer HHI*
Berlin, Germany
behnam.shariati@hhi.fraunhofer.de

Johannes Karl Fischer
*Photonic Networks and Systems*
*Fraunhofer HHI*
Berlin, Germany
johannes.fischer@hhi.fraunhofer.de

*Abstract*—We present a privacy-preserving distributed learning framework for telecom ecosystems in the 6G-era that enables the vision of shared ownership and governance of ML models, while protecting the privacy of the data owners. We demonstrate its benefits by applying it to the use-case of Quality of Transmission (QoT) estimation in multi-domain multi-vendor optical networks, where no data of individual domains is shared with the network management system (NMS).

*Keywords—machine learning, multi-vendor and dis-aggregated networks, distributed learning, shared governance and ownership*

## I. Introduction

The telecom ecosystem is experiencing a significant transformation where Machine Learning (ML) based solutions are expected to play a significant role. On the one hand, the development of sophisticated ML based solutions requires real, field-collected data from the telecom infrastructure. On the other hand, telecom infrastructure is becoming a complicated multi-vendor, multi-tenant, and disaggregated ecosystem, where the availability of telemetry data across the whole ecosystem is not only a technical issue, but also a regulatory one due to data confidentiality and the presence of many players often having conflicts of interests. The regulatory issues on network data sharing and trading impose restrictive measures on the interaction among network data owners (e.g., telecom operators) and ML solution developers (e.g. vendors and research centres). This issue will exacerbate with the introduction of contextual-awareness in beyond 5G and 6G ecosystems where data, which is collected from billions of devices and network users can be of great value for efficient network operation [1]. Therefore, it is crucial to develop a trust-building tool, primarily in the data sharing and trading context, to allow involved parties to collaboratively work on ML model training and validation while ensuring that their privacy and benefits are not compromised.

Privacy-Preserving Artificial Intelligence (PPAI) and one of its most prominent branches, Federated Learning (FL) have been recently proposed as invaluable tools to enable collaborative training of ML models over geographically distributed datasets [2],[3]. PPAI allows different business entities to use models and/or privacy-sensitive data of different owners without compromising their privacy. In other words, these model and data owners interact with one another without necessarily trusting each other. PPAI addresses privacy based on methods such as Differential Privacy (DP) [4], and Secure Multi-Party Computation (SMPC) [5] to realize shared ownership and governance of data and/or ML models.

In this paper, we present a visionary privacy-preserving distributed learning framework, which aims to address this issue by providing a platform for shared governance and ownership of ML models. We report proof-of-concept results in the context of multi-vendor multi-domain disaggregated optical networks in which the Domain Manager (DM) of three different vendors is engaged in shared ML model training process without revealing their data to the operator's NMS.

## II. Architecture of the Distributed Learning Framework

The distributed learning framework trains a global model using data hosted on a set of geo-distributed edge nodes. The proposed solution is based on the work presented in [6] composed of two main components (see Fig.1a), which together contribute to the training of a global ML model using data hosted on the distributed edge nodes. On one side, there is a Training Coordinator Node (TCN) and on the other side, there are several Edge Contributor Nodes (ECNs). The TCN acts as a moderator that manages the overall training procedure. In order for the TCN and ECNs to communicate, a secure communication protocol based on WebSocketSecure (WSS) is adapted [7].

In order to realize a FL architecture, we use the adapted Stochastic Gradient Descent (SGD) [8] algorithm. A typical implementation of this so-called Federated Averaging [3] is presented in Table 1 with a fixed learning rate of η. Each ECN computes the average gradient on its local data on the current

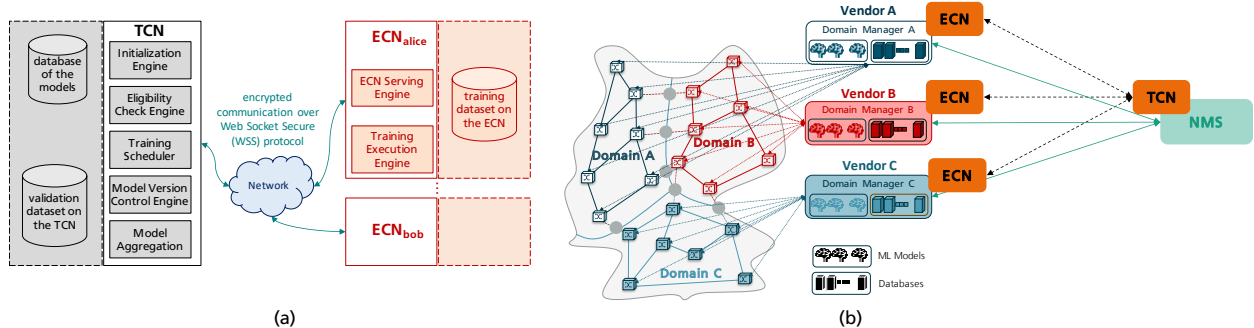

Fig. 1. (a) Modular architrecture of the distributed learning framework, (b) multi-vendor multi-domain telecom ecosystem

**Table 1:** Federated Averaging algorithm. There are K ECNs each of which is indexed by k. B is the set of data batches on an ECN. E is the number of training iterations (epochs) on an ECN, with learning rate η.

**TCN executes**:
  initialize $\omega_0$
  **for** each round $t = 1,2,…$ **do**
    **for** each ECN $k$ **in parallel do**
      $\omega^k_{t+1} \leftarrow ECNupdate(k,\omega_t)$
    $\omega_{t+1} \leftarrow \Sigma^K_{k=1}[(n_k/n) \times \omega^k_{t+1}]$

**ECNupdate**$(k,\omega)$:
  **for** each local epoch i from 1 to $E$ **do**
    **for** batch $b \in B$ **do**
      $\omega \leftarrow \omega - \eta \cdot \nabla l(\omega;b)$
  **return** $\omega$ to server

**Table 2:** Comparison of centralized and distributed learning

| Scenario | Accuracy |
|---|---|
| Shared QoT Model – Centralized | 89.89% |
| Shared QoT Model – Distributed | 89.31% |

model, and the TCN is responsible for the gradient aggregation and update of the global model parameters.

The training workflow comprises four general stages: (1) checking the eligibility of the ECNs, (2) distributing the training configuration among ECNs, (3) reporting the locally obtained models to the TCN, and (4) updating the local models with the newly obtained global model.

## III. RESULTS

In order to show the benefits of our proposed distributed learning framework [9], we present a use-case of Quality of Transmission (QoT) estimation [10] in a multi-domain multi-vendor scenario to showcase the benefit of network data sharing based on mutual trust for network automation.. In this regard, we consider a three-domain optical network where each domain has its own Domain Manager (DM) (see Fig.1b) that hosts the Traffic Engineering Database (TED) of the corresponding domain.

We use our in-house optical network planning tool (see the workflow in Fig.2a) to generate datasets. We perform simulations based on the topology of the network CORONET CONUS (see Fig.2b). We consider 96 equally spaced wavelength channels in C-band with a channel spacing of 37.5 GHz on standard single mode fibre. We then run 16 rounds of simulations and choose 35,216 samples, which are well balanced between the True and False classes of the QoT metric and uniformly distributed over three domain-specific datasets. The used metric, as threshold is set to a BER before forward error correction equal to $3.8 \times 10^{-3}$. We use an Artificial Neural Network with 71 input neurons (equivalent to the used 71 features, which include one-hot encoded ones), 1 hidden layer with 3072 neurons, and two outputs for the binary classifier. We consider two scenarios: (*centralized*), in which we move all the data to the NMS, and (*distributed*) in which we keep the data on the DMs and instead use the distributed learning framework. The results presented in Table 2 show that our proposed framework could obtain a shared ML model for QoT estimation, while keeping the data of each DM on their own site and protect their privacy. The obtained result is comparable with the option where we move all the data to a single location.

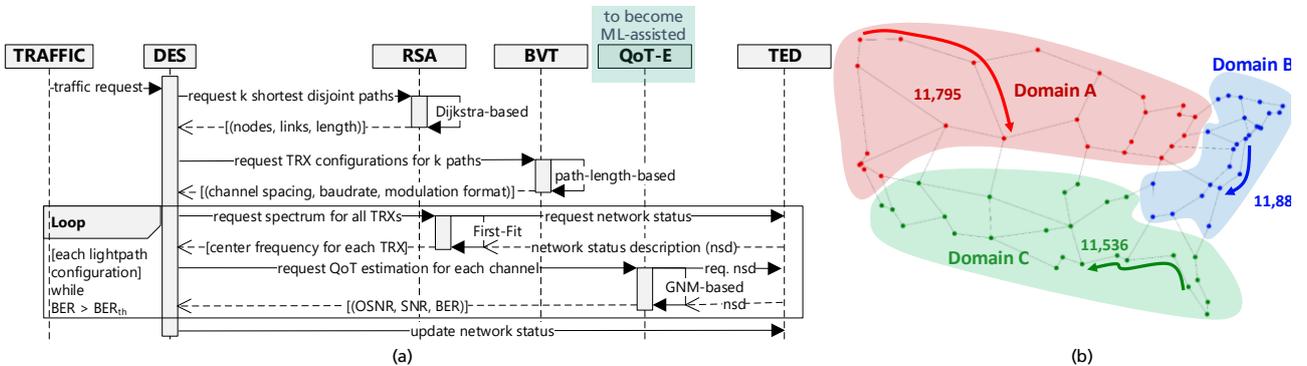

Fig. 2. (a) workflow of the HHI optical network planning suite, (b) multi-domain version of the CORONET CONUS optical network topology. The provided values on each domain represents the number of samples conidered in the domain-specific datasets. DES: Discrete Event Simulator, RSA: Routing and Spectrum Allocation, BVT: Bandwidth Variable Transceiver, QoT-E: Quality of Transmission Estimator, TED: Traffic Engineering Database.